# AUTOMATED VERIFICATION OF ROLE-BASED ACCESS CONTROL POLICIES CONSTRAINTS USING PROVER9

Khair Eddin Sabri

Computer Science Department
King Abdullah II School for Information Technology
The University of Jordan, Amman, Jordan

## ABSTRACT

*Access control policies are used to restrict access to sensitive records for authorized users only. One approach for specifying policies is using role based access control (RBAC) where authorization is given to roles instead of users. Users are assigned to roles such that each user can access all the records that are allowed to his/her role. RBAC has a great interest because of its flexibility. One issue in RBAC is dealing with constraints. Usually, policies should satisfy pre-defined constraints as for example separation of duty (SOD) which states that users are not allowed to play two conflicting roles. Verifying the satisfiability of constraints based on policies is time consuming and may lead to errors. Therefore, an automated verification is essential.*

*In this paper, we propose a theory for specifying policies and constraints in first order logic. Furthermore, we present a comprehensive list of constraints. We identity constraints based on the relation between users and roles, between roles and permission on records, between users and permission on records, and between users, roles, and permission on records. Then, we use a general purpose theorem prover tool called Prover9 for proving the satisfaction of constraints.*

## 1. INTRODUCTION

Information is a valuable asset for many organizations which should be protected from unauthenticated access. The confidentiality property prevents access to or the disclosure of information to unauthorized users. Two main approaches can be used to achieve the confidentiality of information which are cryptography and access control. Cryptography can be used to encrypt the information such that only the authorized users have the required keys to decrypt and reveal the informational. The other approach stores the information behind a trusted server that reveals information to authorized users only. An authentication process is required for this approach.

Several access control methods are proposed in the literature such as mandatory access control (MAC), discretionary access control (DAC), and Role Based Access Control (RBAC). RBAC gives abstraction in dealing with access control such that instead of granting access to individuals it grants access to roles such as instructor, medical doctor, student etc., and then, users are assigned roles. For example, Mark can be assigned the role of instructor. Therefore, Mark is authorized to access all the pieces on information that are granted to the role instructor. Later, if the role of Mark is changed, only the assignment of Mark to the role instructor should be changed, no permission policies are required to be changed. This gives simplicity in dealing with access control policies. Because of its flexibility and suitability for many organizations, RBAC is widely used by many organizations and has attracted many researchers to analyze and extend its applications [1,2,3].

Prior to the deployment of policies, several constraints are needed to be verified such as the separation of duty (SOD) constraint. For example, a record should be partitioned between several users to accomplish a task or users should not be assigned to conflicting roles e.g., a user cannot play the role of instructor and student at the same time. Other constraints may state that each user should be assigned to at least one role, every role has at least one user assigned to it, or a record should not be shared between roles etc. These constraints are important for many organizations to ensure security properties in their systems. The manual management of policies and constraints is time consuming and can lead to errors especially in large systems. Therefore, we need an automated tool for verifying the satisfiability of constraints. Formal methods are suitable for such verification because it gives a precise interpretation of policies and constraints.

In this paper, we develop a theory to specify policies and constrains based on first order logic. Then, we use a theorem prover called Prover9/Mace4 [4] to verify constraints. We use first order logic as it is suitable for such specification and it is a well-known logic which helps in understanding and using our theory. Furthermore, we give a comprehensive list of constraints that can be imposed on the relations in role based access control. We use Prover9 and Mace4 tools to verify the satisfaction of constraints. Beside the ability to specify first order logic, Prover9 and Mace4 have other advantages such as the ability to prove or disprove theorems. Therefore, if we are not able to prove a constraint we can use Mace4 to find a counterexample. Furthermore, Prover9 and Mace4 tools are fully automated. They do not require human interactions and therefore, it will be useful to use them when building a tool that plays as a connecting layer between users and theorem prover.

The rest of this paper is organized as follows. Section 2 summarizes the basic components of RBAC. Section 3 introduces Prover9 and Mace4 tools. Section 4 presents our theory of specifying policies and constraints. Also, it presents the use of Prover9 and Mace4 to prove or disprove the satisfaction of constraints. Furthermore, this section includes our discussion on the use of Prover9 and Mace4. Section 5 presents related work. In Section 6, we conclude and highlight future work.

## 2. ROLE BASED ACCESS CONTROL

A role based access control is an abstraction of mandatory access control. It gives permission to roles instead of users. Users are then assigned to roles such that a user inherits all the permissions assigned to that role. A user can be assigned to more than one role and each role can have many users assigned to it. Role based access control consists mainly of the following components:

- *R*: Set of roles.
- *U*: Set of users.
- *O*: Set of objects called records in this paper.
- *Has_Role* ⊆ *U* × *R* : This relation relates users to their roles.
- *Permission* ⊆ *R* × *O*: This relation relates roles to their allowed records.
- *Has_Access* ⊆ *U* × *O*: This relation relates users to their allowed records. This relation is defined based on the above two relations.

The basic model of RBAC is given in Figure 1. As shown in the figure, there is no direct connection between users and records. Access is made through roles. Usually several constraints are imposed on these three relations. For example, on *Has_Role* relation there are constraints such as every user is assigned to at least one role and each role has at least one user. Another constraint is that a user is not allowed to be assigned to two conflicting roles. Similarly, we have constraints on the other relations. More details are given in the next section.

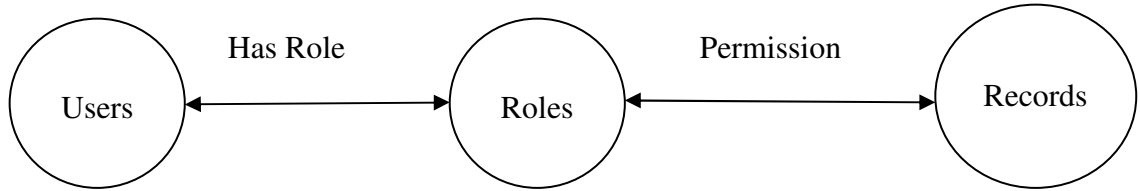

Figure 1: A basic model of RBAC model

## 3. PROVER9 AND MACE4

Prover9 is an automated theorem prover for proving theorems based on first-order and equational logic. Mace4 is an automated tool for proving that the theorem is not true through searching for finite models and counterexamples. Mace4 can help avoid wasting time searching for a proof with Prover9 by first finding a counterexample or by first helping to debug logical specifications. The input file of Prover9 contains the logical specification and a list of goals to be proved. Prover9 handles unsorted terms which can be a variable, a constant, or an *n*-ary function symbol applied to *n* terms. An automatic formula is an *n*-ary predicate symbol applied to *n* terms. A formula can be defined recursively as follows. An atomic formula is a formula. If `F` and `G` are formulas, then `(-F), (F | G), (F & G), (F -> G), (F <-> G)` are formulas. If `F` is a formula and `x` is a variable, then `(all x F)` and `(exists x F)` are formulas. Prover9 distinguishes variables from constants as variables start with lower case *u* through *z*. The meaning of the main symbols of Prover9 is given in Table 1.

1: The main operators of Prover9

| Negation | ~ |
|---|---|
| Disjunction | \| |
| Conjunction | & |
| Implication | -> |
| Equivalence | <-> |
| Universal quantification | all |
| Existential quantification | exists |
| Equality | = |
| Negated equality | != |

## 4. FORMAL SPECIFICATION

To analyze role based access control policies. First, we define lists of roles, users, and objects. Then, we define a relation called *Has_Role* that relates users to roles. Also, we define a relation called *Permission* that relates roles to the allowed objects (e.g., records). As a case study, we present an example related to university. The university has the following users: Sam, David, Mary, John, and James. Each user can play the role of *Student*, *Instructor*, *TA*, *Chair*, *Dean* and *Secretary*. Assume that Sam plays the role of student, David plays the role of TA and instructor, Mary plays the role of secretary, John plays the role of instructor and dean, and James plays the role of chair. We give the specification in Prover9 as:

Has_Role(Sam,Student).
Has_Role(David,TA).
Has_Role(Mary,Secretary).
Has_Role(John,Instructor).
Has_Role(David,Instructor).
Has_Role(James,Chair).
Has_Role(John,Dean).

### 4.1 Constraints on the relation between Role and Users

In this section, we present several constraints that restrict the relation between roles and users.

*Constraint 1:* In some applications, we need to verify that a user playing a role is already playing another role. We can call this a pre-request constraint. In our example, a user playing the role of chair should play the role of instructor as well. Similarly, a user playing the role of dean should play the role of instructor as well. The formula that we need to prove is

x=John| x=Mary -> Has_Role(x,Instructor).

In this formula, we need to explicitly give values for $x$ that represent those users playing the role of dean or chair. Manual writing of this formula is time consuming and can lead to error. However, this step can be automated when a higher layer tool is used that can save all the information in a database where extracting all the names that are playing the role of dean or chair is simple. Similarly, we need to prove that each user playing the role of TA should play the role of student.

x=David -> Has_Role(x,Student).

Both formulas cannot be proved because in the first formula James does not play the instructor role and in the second formula David does not play the role of student. One can by default states that any TA is automatically a student without explicitly specifying that. Similarly, a dean or a chair is also instructor. Therefore, we add into the specification three predicates. The first one indicates that all TAs are students. The second predicate states that all chairs are also instructors. The last predicate states that all deans are instructors. These three predicates are formulated in Prover9 as:

all x (Has_Role(x,TA) -> Has_Role(x,Student)).
all x (Has_Role(x,Chair) -> Has_Role(x,Instructor)).
all x (Has_Role(x,Dean) -> Has_Role(x,Instructor)).

*Constraint 2:* No user can be assigned to two conflicting roles. For example, a student can be a TA but not an instructor. An instructor can be a chair but a secretary cannot be a chair and so on. Therefore, we define the following list of roles to be distinguished: Student, Instructor, and Secretary. We have not added TA, chair, and dean roles to this list because it is specified in Constraint 1 that each chair or dean is an instructor and every TA is a student. To verify Constraint 2, we need to define explicitly that the student, instructor, and secretary roles are distinct. We define that in prover9 as in [5]:

distinct([x, y : z]) ->
( (x != y) &
distinct([x : z]) &
distinct([y : z])).
distinct([Instructor,Secretary,Student]).

The verification of this constraint is achieved by proving the following formula which states the existence of a user playing two distinct roles. Proving this formula indicates that the constraint is not satisfied. Otherwise, indicates that the constraint is satisfied.

exists x exists y exists z(Has_Role(x,y) & Has_Role(x,z) & y!=z).

By running Prover9, a true result is returned stating that a David plays the role of instructor and student. By removing the predicate Has_Role(David,Instructor), Prover9 timed out without finding a result.

*Constraint 3:* Each role must have at least one user. We formulate this property in Prover9 as shown below.

x=TA | x=Student | x=Secretary | x=Chair | x=Instructor | x=Dean   ->   exists y (Has_Role(y,x)).

We are able to prove this constraint, but if we remove the axiom Has_Role(John,Dean) from our specification, we are not able to prove the formula and Mace4 gives a counterexample.

*Constraint 4*:    Similarly, we can verify that each user is assigned at least to one role by using the following formula:

x=Sam | x=David | x=Mary | x=John | x=James
   ->   exists y (Has_Role(x,y)).

*Constraint 5:* Users playing a role $r_1$ can play the role $r_2$ or $r_3$ but not both. For example, an instructor can be a chair or a dean but not both. To formulate this constraint, we add two predicates S1 and S2. The first predicate specifies the instructor role and the second predicate S2 specifies the chair and dean roles as follows.

S1(Instructor).
S2(Chair).
S2(Dean).

Then, we need to prove the following formula which states the existence of a user *x* playing the roles *y*, *z*, and *v* such that *y* has the predicate S1 and *z* and *v* have the predicate S2. If the formula is true, then Constraint 4 is not satisfied, otherwise, the constraint is satisfied.

exists x exists y exists z exists v (Has_Role(x,y) & Has_Role(x,z) & Has_Role(x,v) & S1(y) & S2(z) & S2(v) & y!=z & y!=v & z!=v).

*Constraint 6:* No user in a specified set can be assigned to a role R. For example, we can state that the users Mike and John cannot play the role of chair or the role of dean. First, we identity the names of the users using a predicate called U and specify the roles using a predicate called R as follows:

U(Mike).
U(John).
R(Dean).
R(Chair).

We need to prove the formula that states the existence of a user *x* having the predicate U and playing a role *y* having the predicate R. If the formula is true, then Constraint 6 is not satisfied, otherwise, the constraint is satisfied.

exists x exists y (U(x) & R(y) & Has_Role(x,y)).

## 4.2 Constraints on the relation between Roles and Records

In this section, we present constraints on the relation between roles and their permissions to records. We call this predicate *Permission* and it is defined for our example in Prover9 as:

Permission(Student,REC1).
Permission(Student, REC2).
Permission(TA, REC3).
Permission(Instructor, REC4).
Permission(Instructor, REC5).
Permission(Chair, REC6).
Permission(Dean, REC7).
Permission(Secretary, REC3).

*Constraint 7:* Each record is assigned to at least one role. This constraint is specified as in the following formula.

x=REC1 | x=REC2 | x=REC3 | x=REC4 | x=REC5 | x=REC6 | x=REC7
->   exists y (Permission(y,x)).

*Constraint 8*: Each role is given permission to at least one record. This constraint is specified as in the following formula.

x=Student | x=TA | x=Secretary | x=Instructor | x=Chair | x=Dean   ->   exists y (Permission(x,y)).

*Constraint 9:* We can state that at least two roles should be authorized for permission on specific records. For example, in order to submit the grades of students, the instructor and the chair should both sign the grade records. In this example, we specify a constraint that at least two roles should have permission on the record REC4 using the following formula. If the formula is proved true using Prover9 then the constraint is satisfied, otherwise, the constraint is not satisfied.

z=REC4 -> exists x exists y (Permission(x,z) & Permission(y,z) & x!=y).

*Constraint 10:* A record is assigned to exactly one role. To verify this constraint, we need to verify first Constraint 8 to make sure that each record is assigned to a role. Then, we use formula similar the one defined in Constraint 9. If the formula is proved true then the constraint is not satisfied, otherwise, the constraint is satisfied.

*Constraint 11:* A role cannot have permission to conflicting records. To specify this constraint, we need first to specify the conflicting records using the predicate *distinct* specified previously.

distinct([REC1,REC2,REC5]).

Then, we need to prove the following formula. If the formula is proved true then the constraint is not satisfied. Otherwise, the constraint is satisfied.

exists x exists y exists z (Permission(x,y) & Permission(x,z) & y!=z).

## 4.3 Constraints on the relation between Users and Records

We need to link users to permissions on records. Therefore, we add the following predicate into our specification which states that a user *x* playing the role *y* that has permission to access record *z* implies that the user *x* can access the record *z*.

all x all y all z (Has_Role(x,y) & Permission(y,z)   -> Has_Access(x,z)).

*Constraint 12:* Each user is given permission to at least one record. Verifying this constraint is achieved by verifying Constraint 4 and Constraint 8.

*Constraint 13:* At least two users should have an access to a record.

z=REC1 -> exists x exists y (Has_Access(x,z) & Has_Access(y,z) & x!=y).

This formula states that two different users should have access the record REC1. Using the same formula we can verify any number of users. For example, to state that three different users should be able to access the record REC1, we need to prove the formula

z=REC3 -> exists x exists y exists v (Has_Access(x,z) & Has_Access(y,z) &
Has_Access(v,z)   & x!=y & x!=v & y!=v).

## 4.4 Constraints on the Relation between Users, Records, and Roles

In this section, we state a constraint that involves users, roles and permissions to records.

*Constraint 14:* At least three users belonging to two different roles should be able to access the record REC3.

z=REC3 -> exists x exists y exists v exists r1 exists r2 exists r3 (Has_Access(x,z) &
Has_Access(y,z) & Has_Access(v,z) & Has_Role(x,r1) & Has_Role(y,r2) &
Has_Role(v,r3) & (r1!=r2 | r1!=r3)   & x!=y & x!=v & y!=v).

## 4.5 Discussion

We found that first order logic is suitable for specifying policies and constraints. Furthermore, we found that Prover9 is efficient for proving theorems and Mace4 for disproving them. Based on our experience in specifying and proving constraints, we highlight the following two issues.

The first issue is the incompleteness of our specification. Therefore, we face a case in which a theorem is neither proved nor disproved. As an example of such a case, we have in our specification that the role of David is TA and that can be proved easily in Prover9. However, we cannot prove or disprove that David does not play the role of chair because no axioms indicating that. Therefore, to have a complete specification we need to explicitly state that David does not play the role of chair as -Has_Role(David,Chair). This completeness issue directs us on writing our formula for proving the first constraint. To prove that every chair is also an instructor, it could be straight forward to write our formula as all x Has_Role(x,Chair) -> Has_Role(x,Instructor).

Instead of
x=Mary -> Has_Role(x,Instructor).

However, we cannot use the first formula because we do not have axioms indicating whether David is a chair or not. To solve this problem, we have two options: either making our specification complete by stating all negation policies such as David is not a chair or explicitly giving the values of *x* to all users playing the role of chair or dean as we did in our example. This should not be an issue because our theory should be integrated with a high level tool that generates formulas automatically and calls Prover9 to prove theorems.

The second issue is that in some cases as in constraint 5 it is better to write the formula in a negative way such as proving the formula indicates that the constraint is not satisfied while not being able to prove the theorem indicates that the constraint is satisfied. In our example, Prover9 gives results within seconds. However, we are not aware of the time required to find an answer to this constraint on a large scale that contains hundreds or thousands of records which will be our future work.

# 5. LITERATURE REVIEW

Several researchers use Prover9 in their formal analysis. For example, Sabri and Khedri [6,7,8] use Prover9 in verifying security properties at an abstract level in systems that use key assignment schemes. Cunha and Macedo [9] present a framework for unbounded verification of Alloy specifications using the automatic theorem prover Prover9. Rusnok and Adlassnig [10] use Prover9 for detecting inconsistency in health records. Furthermore, Prover9 is used to prove theorems related to algebraic structures [11,12,13]. To the best of our knowledge our work is the first that uses Prover9 in proving RBAC constraints.

Role-Based access control has been investigated by many researchers. Ahn and Sandhu [14] describe a specification language called RCL 2000. They give a formal syntax and semantic for RCL 2000 and demontrate its soundness and completeness. Furthermore, they include session-based dynamic SOD. Ahn and Shin [15] present an approach to specify constraints using a declarative language called Object Constraints Language (OCL) that is part of the Unified Modeling Language (UML). Crampton [16] presents a set-based specification language and discusses the enforcement of constraints. Ray et. al. [17] presents an approach for systematically incorporating RBAC policies into an application design model specified by using UML. Our work is different from the existing ones in using the general purpose tools Prover9 and Mace4 which enable us to prove or disprove theorems. The verification of constraints is fully automated. Furthermore, we present a comprehensive list of constraints and propose ones not discussed in the literature.

# 6. CONCLUSION AND FUTURE WORK

In this paper, we develop a theory for specifying role-based access control policies based on first order logic. Furthermore, we present a comprehensive list of constraints on policies and prove them automatically using the general purposes tool Prover9. We show that all our constraints can be expressed and verified using Prover9. As a future work, we intend to build a tool that acts as an intermediate layer between users and Prover9 tool. This tool enables users, including inexperienced ones, to apply Prover9 on verifying the satisfaction of constraints. Furthermore, we aim at studying the performance of proving constraints in real applications. Finally, we intend to focus on dynamic constraints as it is not discussed in this paper.

# REFERENCES

[1] R. S. Sandhu, E. J. Coyne, H. L. Feinstein and C. E. Youman, "Role-Based Access Control Models," Computer, vol. 29, no. 2, pp. 38--47, 1996.

[2] E. Bertino, P. A. Bonatti and E. Ferrari, "TRBAC: A Temporal Role-based Access Control Model," ACM Transactions on Information and System Security, vol. 4, no. 3, pp. 191-233, 2001.

[3] S. Rizvi, A. Mendelzon, S. Sudarshan and P. Roy, "Extending Query Rewriting Techniques for Fine-grained Access Control," in ACM SIGMOD International Conference on Management of Data, Paris, France, 2004.

[4] W. McCune, Prover9 and Mace4, 2005-2010.

[5] D. A. Wheeler, "Fully Countering Trusting Trust through Diverse Double-Compiling," Fairfax, VA, 2009.

[6] K. E. Sabri, "Algebraic Analysis of Object-Based Key," Journal of Software, vol. 9, no. 9, pp. 2033-2042, 2014.

[7] K. E. Sabri and R. Khedri, "Algebraic Framework for the Specification and Analysis of Cryptographic-Key Distribution," Fundamenta Informaticae, vol. 112, no. 4, pp. 305-335, 2011.

[8] K. E. Sabri and R. Khedri, "A Generic Algebraic Model for the Analysis of Cryptographic-Key Assignment Schemes," in Foundations and Practice of Security, Montreal, QC, Canada, 2013.

[9] A. Cunha and N. Macedo, "Automatic Unbounded Verification of Alloy," CoRR, vol. abs/1209.5773, 2012.

[10] P. Rusnok and K. Adlassnig, "Detection of Inaccuracy in a Medical Knowledge Base Using a Classical Theorem Prover," in Health Informatics Meet Ehealth, Vienna, Austria, 2010.

[11] H.-H. Dang and B. Möller, "Simplifying Pointer Kleene Algebra," in 1st Workshop on Automated Theory Engineering, Poland, 2011.

[12] P. Höfner and S. Georg, "Automated Reasoning in Kleene Algebra," in 21st International Conference on Automated Deduction, Bremen, Germany, 2007.

[13] J. D. Phillips and D. Stanovsk, "Automated Theorem Proving in Loop Theory," Artificial Intelligence Communications , vol. 23, no. 2-3, pp. 267-283, 23/2-3.

[14] G.-J. Ahn and S. Ravi, "Role-based Authorization Constraints Specification," ACM Transactions on Information and System Security, vol. 3, no. 4, pp. 207-226, 2000.

[15] G.-J. Ahn and M. Shin, "Role-based authorization constraints specification using Object Constraint Language," in The Tenth IEEE International Workshops on Enabling Technologies: Infrastructure for Collaborative Enterprises, Cambridge, MA, 2001.

[16] J. Crampton, "Specifying and Enforcing Constraints in Role-based Access Control," in the Eighth ACM Symposium on Access Control Models and Technologies, Como, Italy, 2003.

[17] I. Ray, N. Li, R. France and D.-K. Kim, "Using UML to Visualize Role-based Access Control Constraints," in The Ninth ACM Symposium on Access Control Models and Technologies, Yorktown Heights, New York, USA, 2004.

# AUTHORS

Khair Eddin Sabri has been working as an assistant professor in the Computer Science Department at The University of Jordan since 2010. He obtained his B.Sc. degree in Computer Science from the Applied Science University, Jordan in June 2001. He also received M.Sc. degree in Computer Science from The University of Jordan in January 2004 and a Ph.D. degree in Software Engineering from McMaster University, Ontario Canada in June 2010. He is a member of the Formal Requirements and Information Security Enhancement (FRAISE) Research Group. His main research interest is the formal verification and analysis of security properties.

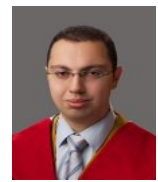